\documentclass[iop,revtex4]{emulateapj}

\usepackage{apjfonts}
\usepackage{times}

\usepackage{amsmath}
\usepackage{graphicx}
\usepackage{natbib}
\usepackage{epstopdf} 
\usepackage{subfigure}
\usepackage{color}
\usepackage{lscape}

 \font\sevenrm=cmr7 scaled 1000

\newcommand{\mbh}{$M_{\rm BH}$}     
     
\newcommand{\ergs}{erg~s$^{\rm -1}$}
       
\newcommand{\CIV}{C~{\sevenrm IV}}
\newcommand{\NV}{N~{\sevenrm V}}

\newcommand{\OIII}{[O~{\sevenrm III}]}

\newcommand{\HeII}{He~{\sevenrm II}}
\newcommand{\FeII}{Fe~{\sevenrm II}}

\newcommand{\Lya}{Ly$\alpha$}

\newcommand{\Hb}{H$\beta$}

\newcommand{\MgII}{Mg~{\sevenrm II}}
\begin{document}
\title{
The \FeII/\MgII\ flux ratio of low-luminosity quasars at z $\sim$3}
\author{Jaejin Shin$^{1}$}
\author{Tohru Nagao$^{2}$}
\author{Jong-Hak Woo$^{1}$\altaffilmark{,3}}
\author{Huynh Anh N. Le$^{1}$}
\affil{
$^1$Astronomy Program, Department of Physics and Astronomy, 
Seoul National University, Seoul, 151-742, Republic of Korea\\
$^2$Research Center for Space and Cosmic Evolution, Ehime University, 
Bunkyo-cho 2-5, Matsuyama, Ehime 790-8577, Japan
}

\altaffiltext{3}{Author to whom any correspondence should be addressed}

\begin{abstract}
The \FeII/\MgII\ line flux ratio has been used to investigate
the chemical evolution of high redshift active galactic nuclei (AGNs). No strong evolution has been found out to $z\sim6$, implying
that the type 1a supernova activity has been already occurred in early universe.
However, the trend of no evolution can be caused by the sample selection bias since the previous studies have utilized mostly very luminous AGNs, which
may be already chemically matured at the observed redshift. 
As motivated by the previously reported correlation between AGN luminosity and metallicity, 
we investigate the \FeII/\MgII\ flux ratio over a large dynamic range of luminosity, by adding 
a new sample of 12 quasars at $z\sim3$, of which the lower luminosity limit is more than 1 dex smaller than that of the previously studied high-z quasars.
Based on the Gemini/GNIRS observations, we find that the seven low-luminosity quasars with a mean bolometric luminosity log $L_{\rm bol}\sim46.5 \pm 0.2$ has an average \FeII/\MgII\ ratio of $0.68\pm0.11$ dex. This ratio is comparable to that of high-luminosity quasars (log $L_{\rm bol}\sim 47.3 \pm 0.3$) in our sample (i.e., \FeII/\MgII\ ratio of $0.59\pm0.15$ dex) and that of the previously studied high-luminosity quasars at higher redshifts. 
One possible scenario is that the low-luminosity quasars in our sample are still relatively luminous and already chemically matured. 
 To search for chemically-young AGNs, and fully understand the chemical evolution based on the \FeII/\MgII/ flux ratio,
it is required to investigate much lower-luminosity AGNs.
\end{abstract}

\section{INTRODUCTION} \label{section:intro}
Chemical properties of galaxies are powerful diagnostics in understanding galaxy evolution
since the chemical composition is determined by the star-formation history 
as well as the gas inflow and outflow. As the metallicity measurements of galaxies
at high redshifts are particularly interesting for exploring the early phase of galaxy evolution,
the gas phase metallicity of star-forming galaxies has been measured even at $z>3$ \citep[e.g.,][]{Maiolino2008,Mannucci2009,Nagao2012,Amorin2014,Troncoso2014,
Sanders2016,Onodera2016,Shapley2017}. 
However, such metallicity measurements are generally very challenging due to the faintness of star-forming galaxies
at high-z. Alternatively active galactic nuclei
(AGNs) powered by a supermassive black hole (SMBH) have been utilized for 
inferring the chemical composition of galaxies in the early Universe, owing to their high apparent magnitude
as well as numerous metallic emission lines seen in the rest-frame ultraviolet (UV) spectra. 

It has been pointed out through theoretical simulations that various combinations of broad
emission lines observed in the UV spectra of quasars (especially with the \NV$\lambda$1240 line,
i.e., \NV$\lambda$1240/\CIV$\lambda$1549 and \NV$\lambda$1240/\HeII$\lambda$1640;
hereafter \NV/\CIV\ and \NV/\HeII) are useful to infer the gas metallicity of broad-line regions
(BLRs) in quasars \citep[e.g.,][]{Hamann1992,Hamann1993,Nagao2006}. By using these flux
ratios, a positive correlation between AGN luminosity and the BLR metallicity has been 
reported \citep[e.g.,][]{Hamann1993,Shemmer2002,Shemmer2004,Warner2004,Nagao2006,
Shin2013}. This correlation can be interpreted as the result of the underlying correlation 
between the mass of SMBHs ($M_{\rm BH}$) and the BLR metallicity 
\citep{Warner2003,Warner2004,Matsuoka2011b}, at least in the high-redshift Universe as AGN 
luminosity roughly scales with \mbh\ (but see also \citealt{Shin2013} for low-z AGNs). 
Since $M_{\rm BH}$ is tightly correlated with the mass of the 
spheroidal component of galaxies \citep[e.g.,][]{Magorrian1998,Ferrarese2000,
Gebhardt2000,Marconi2003,Haring2004,Kormendy2013,Woo2013,Woo2015}, it is possible 
to interprete that the AGN luminosity-BLR metallicity correlation in quasars is caused by the mass-metallicity relation of galaxies 
\citep[e.g.,][]{Lequeux1979,Tremonti2004}. In other words, more massive galaxies with higher gas metallicity 
host more massive black holes, which are observed as on-average higher-luminosity AGNs, if we assume the Eddington ratio distribution is similar, regardless of the mass scale.

Surprisingly, studies of the BLR metallicity based on \NV/\CIV\ using quasars over a wide 
range of redshift have found no metallicity evolution out to $z \sim 4$ or even higher redshift (e.g., \citealt{
Nagao2006,Matsuoka2011b,Xu2018}
; see also \citealt{Juarez2009,DeRosa2014,Mazzucchelli2017}). Note that the 
gas metallicity of narrow-line regions (NLRs) also shows no redshift evolution \citep[e.g.,][]
{Nagao2006b,Matsuoka2009,Matsuoka2011a,Dors2014}. These results are unexpected,
because the mass-metallicity relation of star-forming galaxies shows a significant redshift
evolution toward $z \sim 3$, even for the most massive galaxies  (e.g., \citealt{Maiolino2008,
Troncoso2014,Onodera2016}; see also \citealt{Sommariva2012}). 

In investigating the chemical evolution of AGNs, the flux ratio of the UV \FeII\ multiplet emission to the 
\MgII$\lambda$2800 emission (hereafter \FeII/\MgII) measured from the UV spectra of quasars provides a more 
interesting diagnostic, since the timescale of the iron enrichment is much longer than other metallic elements. 
This is because the iron is mainly ejected from the type Ia super novae (SNIa) while the 
$\alpha$-elements including magnesium are produced mostly by the core-collapse super novae.
Thus, the enrichment timescale of magnesium is much shorter than that of iron. 
Various attempts have been made to identify the redshift where a significant decrease of 
\FeII/\MgII\ flux ratio can be detected due to the decrease of the iron abundance relative to magnesium. 
However, no significant redshift evolution of \FeII/\MgII\ flux ratio has been observed out to $z \sim 7$ \citep[e.g.,][]{Barth2003,
Dietrich2003,Maiolino2003,Iwamuro2004,Jiang2007,Kurk2007,DeRosa2011,Mazzucchelli2017}.
One potential explanation of this mystery is the sample bias. 
The previous studies have used mainly high-luminosity quasars (i.e., $L_{\rm bol} \gtrsim10^{47}$ \ergs) at high 
redshifts for investigating the \FeII/\MgII\ ratio evolution. 
Such very luminous quasars, with presumably high mass black holes, are likely to be already chemically matured, 
as expected from the mass-dependent evolution of galaxies (\citealt{Juarez2009}; see also \citealt{Kawakatu2003}).

To explore potential candidates of chemically young quasars, we focus on low-luminosity quasars, 
that have not been extensively investigated. By measuring the \FeII/\MgII\ flux ratio of low-luminosity 
quasars, we aim at shedding light on the chemical evolution of quasars in the early Universe. 
In \S2, we describe the sample and observations. The analysis and results follow in 
\S3 and \S4. In \S5, we provide the discussion based on the results. We adopt a cosmology of 
$H_{\rm 0}= 70$ km  s$^{-1}$ Mpc$^{-1}$,  $\Omega_{\Lambda}=0.7$ and 
$\Omega_{\rm m}=0.3$.

\begin{deluxetable*}{cccccccc} 
\tablewidth{0pt}
\tablecolumns{9}
\tabletypesize{\scriptsize}
\tablecaption{Log of observations}
\tablehead{
\colhead{Target} &
\colhead{RA} &
\colhead{Dec} &
\colhead{Redshift} &
\colhead{$m_{I}$} &
\colhead{Observation Date} &
\colhead{Exposure} &
\colhead{Standard star}
\\
&
\colhead{(deg)} &
\colhead{(deg)} &
&
\colhead{(mag)} &
&
\colhead{(sec)} &
\\
\colhead{(1)} &
\colhead{(2)} &
\colhead{(3)} &
\colhead{(4)} &
\colhead{(5)} &
\colhead{(6)} &
\colhead{(7)}  &
\colhead{(8)} 
}
\startdata
SDSS J082854.44+571637.2	&$	127.22686	$&$	57.27702	$&$	3.383	$&$	20.20	$&	2012 Feb 09	&$	300\times32	$ & HIP 42124 (A5V)\\
SDSS J124652.80+545140.6	&$	191.72000	$&$	54.86130	$&$	3.360	$&$	20.11	$&	2012 Feb 09	&$	300\times24	$ & HIP 61510 (F0V)\\
SDSS J120308.69+552245.8	&$	180.78622	$&$	55.37941	$&$	3.355	$&$	20.15	$&	2012 Feb 09	&$	300\times16	$ & HIP 63153 (F4V)\\
SDSS J081528.12+344737.0	&$	123.86720	$&$	34.79362	$&$	3.200	$&$	20.19	$&	2015 Mar 19	&$	140\times36	$ & HIP 47631 (F3V)\\
SDSS J095617.14+373224.7	&$	149.07143	$&$	37.54022	$&$	3.245	$&$	20.38	$&	2015 Mar 19	&$	140\times36	$ & HIP 51914 (F3V)\\
SDSS J133600.20+391826.2	&$	204.00087	$&$	39.30729	$&$	3.228	$&$	20.40	$&	2015 Mar 20	&$	140\times36	$ & HIP 72469 (F2V)\\ 
SDSS J142903.86+062620.4	&$	217.26610	$&$	  6.43901	$&$	3.268	$&$	20.49	$&	2015 Mar 22	&$	140\times36	$ & HIP 77946 (F0V)\\
\hline
\hline
SDSS J010049.76+092936.1	&$	15.20736	$&$	9.49337	$&$	3.118	$&$	18.65	$&	2017 Nov 23	&$	300\times8	$&	HIP 06373	(F4V)	\\
SDSS J013735.46--004723.4	&$	24.39779	$&$	-0.78983	$&$	3.209	$&$	19.26	$&	2017 Nov 24	&$	300\times20	$&	HIP 01964	(F3V)	\\
SDSS J093514.41+343659.5	&$	143.81004	$&$	34.61654	$&$	3.227	$&$	18.40	$&	2017 Nov 24	&$	300\times8	$&	HIP 49548	(F3V)	\\
SDSS J223408.99+000001.6	&$	338.53748	$&$	0.00047	$&$	3.026	$&$	17.29	$&	2017 Nov 22	&$	248\times4	$&	HIP 107167	(F2V)	\\
SDSS J231934.77--104037.0	&$	349.89489	$&$	-10.67695	$&$	3.170	$&$	18.13	$&	2017 Nov 23	&$	300\times8	$&	HIP 112782	(F0V)	
\enddata
\label{table:prop}

\tablecomments{
    Col. (4, 5): Redshift and $I$-band model magnitude are taken from SDSS DR12 \citep{Alam2015}. 
    Col. (8): Standard stars used for the telluric absorption correction and the flux calibration. 
    The spectral type is shown in the parenthesis.}
\end{deluxetable*}

\section{Sample and Data} \label{section:sample}
\subsection{Sample selection and observations}\label{section: RD}
To investigate the \FeII/\MgII\ ratio of low-luminosity quasars at redshift z$\sim$3, we selected seven 
quasars from the Sloan Digital Sky Survey (SDSS; \citealt{York2000}) Data Release 7 (DR7; 
\citealt{Abazajian2009}) quasar catalog \citep{Schneider2010} and Data Release 9 (DR9; 
\citealt{Ahn2012}) quasar catalog \citep{Paris2012}, with the $I$-band model magnitude in the
range of 20.0--20.5, which is much fainter ($\sim$1 dex) than the sample of previous studies 
\citep[i.e.,][hearafter D03 and M03]{Dietrich2003,Maiolino2003}. We limited the redshift 
range of the sample within $z= 3.0-3.5$, in order to cover a wide-wavelength range 
(i.e., 2200--3090\AA\ in the rest frame) of the \FeII\ multiplet and AGN continuum emission, 
and to avoid strong atmospheric absorptions, 
in the observed spectra. 
We conducted our observations with Gemini 
Near-Infrared Spectroscopy (GNIRS; \citealt{Elias2006}) on the Gemini North telescope, 
of which the cross-dispersed mode covers the wavelength range of $\sim$0.9--2.5 
$\mu$m simultaneously. We observed three quasars on 2012 Feb. 9 (ID: 2012A-C-003, 
PI: T. Nagao) and four quasars on 2015 Mar. 19--22 (ID: 2015A-Q-203, PI. J. Shin). 
Additionally, we observed five high-luminosity quasars ($I$-band magnitude range of 
17.0--19.0) on 2017 Nov 22--24 (ID: 2017B-Q-53, PI. J. Shin) as a comparison sample. 
We used the short camera (0.15\arcsec/pixel) and the grating of 31.7 l/mm along with a 
0.675\arcsec-width slit by considering the typical seeing size at the Gemini North site 
($\sim$0.6\arcsec\ for 70\%-ile), resulting in the spectral resolution of $R \sim 700$. The 
typical seeing size in our three observing runs was $\sim 0.7 \arcsec$, $\sim 0.5 \arcsec$, 
and $\sim 0.5 \arcsec$, respectively. The data were obtained using so-called ABBA-pattern nodding. 
We present the basic information of our observations in Table 1.   

In addition, we selected low-redshift quasars from the quasar property catalog of the SDSS DR7 
quasars \citep{Shen2011}, in order to investigate the \FeII/\MgII\ of low-redshift quasars. We initially 
selected 4,351 quasars at $0.75<z<1.96$ with signal-to-noise ratio of $S/N>20$ at 3000\AA. 
Then, we excluded 248 objects, which show strong absorption lines or weak \MgII\ line (i.e., amplitude-to-noise ratio is less than 3).
Thus, the final sample of 4,103 low-redshift quasars is used for comparison. 

\subsection{Data reduction for Gemini/GNIRS dataset}\label{section: RD}

The data were reduced by using IRAF Gemini/GNIRS package \citep{Cooke2005}. We 
followed the standard data reduction method, including the flat fielding, sky subtraction, 
wavelength calibration, spectrum extraction, and flux calibration.
We extracted the one-dimensional spectrum by adopting an aperture size of 5 pixels 
($\sim$0.75\arcsec). Then we corrected for the telluric absorption features and calibrated the 
flux, based on the spectra of standard stars listed in Table 1. We checked that our flux-calibrated 
spectra are consistent with the SDSS spectra at $\sim9000$\AA\ typically within $\sim10\%$.

\section{Analysis}\label{section:analysis}
In this section, we present the procedure of our fitting analysis for measuring spectral features, 
that are sensitive to chemical properties of BLRs. Specifically, we describe the fitting 
strategy for the spectral regions around \MgII, \Hb, and \CIV, in the subsequent subsections. 
We note that we could not fit \Hb\ and \CIV\ for our low-redshift comparison sample, since 
their SDSS spectra only cover the \MgII\ region.
During the fitting process, we adopted the redshift given in the SDSS database (see Table 1)
as an initial parameter. However, the center of each spectral feature was allowed to change during
the fitting, since emission lines from BLRs sometimes show a velocity shift due to outflows 
\citep[e.g.,][]{Shin2017}.
Note that in this work we do not investigate any kinematical properties of emission lines since 
our immediate interest is the flux ratios of broad emission lines. 

\subsection{\MgII\ and \FeII}\label{section: RD}

For investigating the \FeII/\MgII\ flux ratio, we need to measure the flux
of the \MgII\ and \FeII\ emission lines. Since the \MgII\ is blended with the \FeII\ multiplet, we conducted 
multi-component fitting with \MgII, \FeII, and continuum components (power-law continuum 
and Balmer pseudo-continuum) rather than fitting the \MgII\ and \FeII\ lines separately, 
to minimize measurement uncertainties.

\subsubsection{Continuum components}
First of all, we considered two components in the continuum model: power-law component 
($F_{\lambda}\propto\lambda^{\alpha}$), and Balmer pseudo continuum \citep{Grandi1982}, which 
is expressed as
\begin{equation}
  F_{\lambda}^{\rm BaC} = 
       F^{\rm BE} B_{\lambda}(T_{\rm e})
       (1-e^{- \tau_{\lambda}(\frac{\lambda}{\lambda_{\rm BE}})^{3}})
\end{equation}
where $B_{\lambda}(T_{\rm e})$ is the Planck function at the electron temperature $(T_{\rm e})$. 
$\tau_{\lambda}$ is the optical depth at the Balmer edge; $\lambda_{\rm BE}=3646$\AA. 
$F^{\rm BE}$ is the normalized flux density 
which is normally determined at 
$\lambda_{\rm BE}=3675$\AA\ where the \FeII\ emission is not present. 
We used the same parameters (i.e., $T_{\rm e}=15,000$~K, $\tau_{\lambda}=1$, and
$F^{\rm BE}$=0.3$\times$ $F^{\rm power-law}_{3675 \AA\ }$) as adopted in the previous studies
\citep{Dietrich2003, Kurk2007, DeRosa2011} to avoid systematic difference in comparing with previous results.

\subsubsection{MgII}
In order to fit the line profile of \MgII\, we adopted a single Gaussian model.
We note that the single Gaussian profile usually does not perfectly reproduce the MgII line
profile due to its intrinsically complex velocity profile. Instead, multiple Gaussian (double or triple) or the 
Gauss-Hermite \citep{Van1993} series has been used for the \MgII\ line fitting 
\citep[e.g.,][]{McLure2002,McGill2008,Shen2011,Karouzos2015}. For our sample of high-z
quasars, the Gauss-Hermite series tends to fit the noise in the spectra due to the relatively low S/N. 
Thus, we simply adopted a single Gaussian profile. We also used a single 
Gaussian model for the comparison sample of low-z quasar for consistency. 
When we adopted a Gauss-Hermite model, the measured \FeII/\MgII\ flux ratio is on average 
7\%\ higher than the measurement based on the single Gaussian, suggesting that the choice of the
fitting model does not significantly changes the results described in the following sections.

\subsubsection{Iron templates}
The \FeII\ multiplet emission seen in the rest-frame UV spectra is often fitted with an iron 
template. 
Previous studies on the \FeII/\MgII\ flux ratio commonly used two iron templates, respectively, 
provided by \citet[][hearafter VW01]{Vestergaard2001} and \citet[][hearafter T06]{Tsuzuki2006}
\citep[e.g.,][]{Maiolino2003,Dietrich2003,Kurk2007,DeRosa2011,Dong2011,Sameshima2017}. 
However, the VW01 template has no information of iron around \MgII\ (2770-2820\AA) 
since the template was made by masking the \MgII\ region based on the spectrum of I Zwicky 1. 
To overcome this issue, \cite{Kurk2007} modified the VW01 template (hereafter the modified 
VW01 template) by adding a constant flux density, which is calculated as the average flux density at 
2930--2970\AA, to the \MgII\ window (2770--2820\AA). On the other hand, T06 constructed a 
new iron template, which covers the \MgII\ region, using a photoionization modeling 
\citep[Cloudy;][]{Ferland1998}. Using the two templates by VW01 and T06, \cite{Woo2018} 
compared \MgII\ and \FeII\ fitting results and found that the fitting result with the VW01 template 
tends to underestimate \FeII\ and overestimate \MgII\ fluxes, compared to the measurements using T06. 
To understand the effect of iron template for the \FeII/\MgII\ flux ratio, we adopted three iron templates;
1) the T06 template, 2) the VW01 template, and 3) the modified VW01 template for comparison. 


\subsubsection{Fitting procedure}

In the fitting procedure, the iron template was convolved with a Gaussian kernel with a range of
velocity dispersion 500--10000 km~s$^{-1}$. In addition, to compare with the result of \cite{DeRosa2011}, 
we also adopted the modified VW01 template with a fixed velocity dispersion at 1860 km/s.

Because of the similar ionization energies (15.03 eV for Mg$^+$ and 16.18 eV for Fe$^+$), we 
fixed the velocity shift of \MgII\ and \FeII\ to be the same, and limited the difference in their 
velocity dispersions within 10\%. The \MgII\ and \FeII\ fluxes were then measured from the 
best-fit model. Especially, for the \FeII\ flux, we summed the total \FeII\ flux between 
2200--3090\AA, as conventionally used, for comparing with the previous studies.

\begin{figure}
\includegraphics[width = 0.45\textwidth]{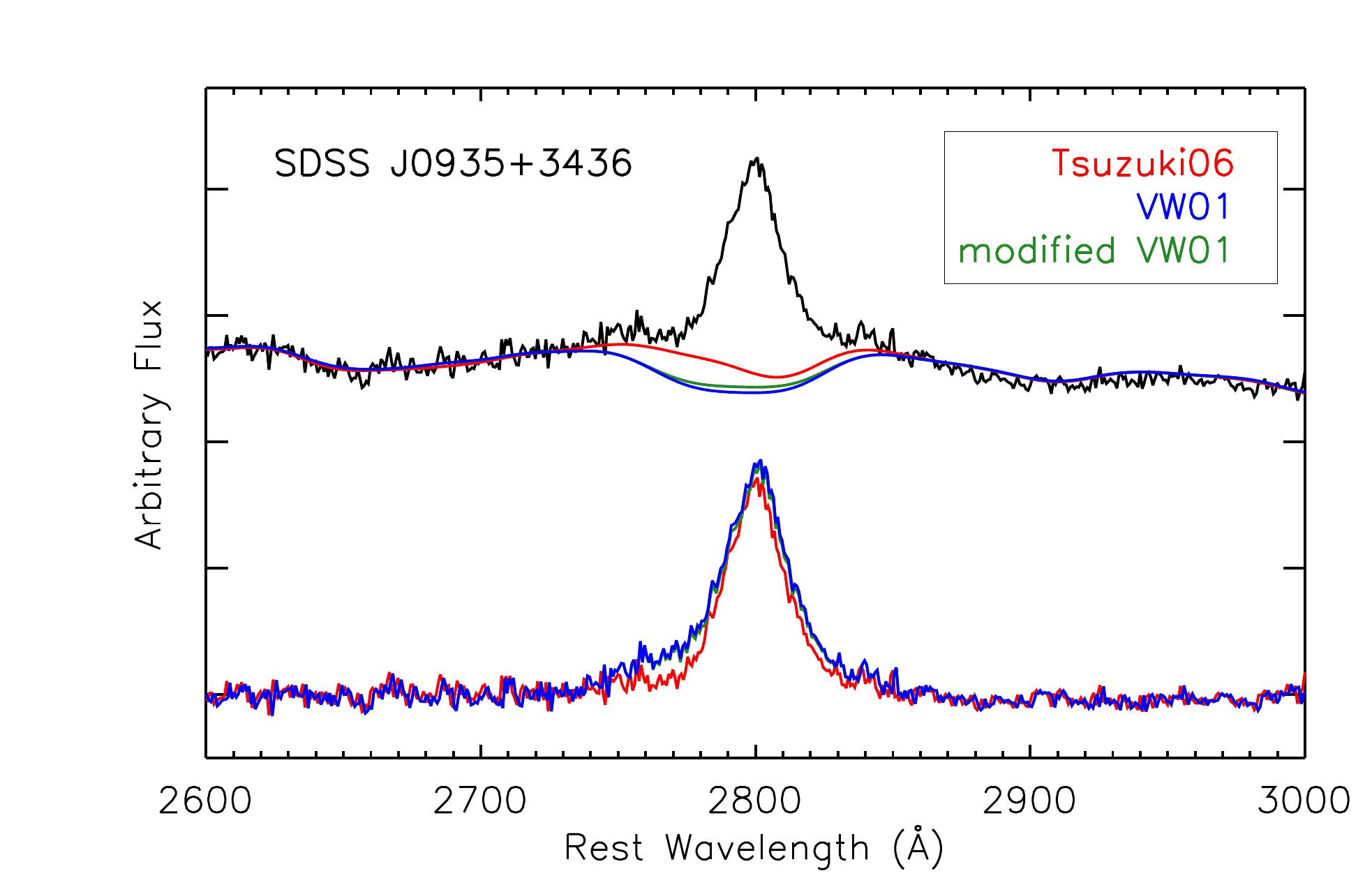}
\caption{
     Fitting results with three iron templates for SDSS J0935+3436. 
     Top: Fully reduced GNIRS spectra in the spectral region around the \MgII\ emission after 3 pixel
     smoothing, where the best-fit of power-law continuum+Balmer pseudo-continuum+\FeII\ are
     overlaid. Color represents each iron template used in fitting procedure (red: the T06 template, 
     blue: the VW01 template, and green: the modified VW01 template). 
     Bottom: Continuum and \FeII\ model subtracted residual which represents \MgII\ line. 
     Colors are the same as in the top panel.
     \label{fig:allspec1}}
\end{figure}

In Figure~1, we present an example of fitting results with three iron templates (i.e., the VW01 template, 
the modified VW01 template, and the T06 template). The VW01 template 
and the modified VW01 template are lacking the \FeII\ component at the wavelength range of the 
broad \MgII\ emission line, resulting in weaker \FeII\ and stronger \MgII\ flux than the fitting 
result with the T06 template. 
Thus, we decide to adopt the measurements based on the T06 template as our main result. 
However, to compare with previous studies, we also use the measurements with the VW01 template. 
We will discuss the effect of iron model for \FeII/\MgII\ flux ratio in \S5.2.

The fitting results for our 12 targets based on the T06 template are shown in Figure 2. We note 
that the data around 3200\AA\ are very noisy for SDSS J0956+3732, SDSS J1336+3918, and 
SDSS J1429+0626, due to the gap between the two orders of GNIRS. However, It does not affect the 
fitting process since they are out of the range of our interest. The measurement error is estimated 
through Monte-Carlo simulations by generating 1,000 mock spectra by randomizing the error of the raw 
spectra. We fitted the 1,000 mock spectra with the same constraints described above and adopted 
the mean and standard deviation of the distribution of the 1,000 measurements as the final
measurement and uncertainty. The derived \FeII/\MgII\ ratio and their uncertainty, respectively with the four 
iron templates, are given in Table 2. We adopted 2 sigma upper limit of \FeII/\MgII\ for 
SDSS J0956+3732 due to the marginal detection of \FeII\ (see Figure 2).

\begin{figure}
\centering
\includegraphics[height=15cm, width = 0.51\textwidth]{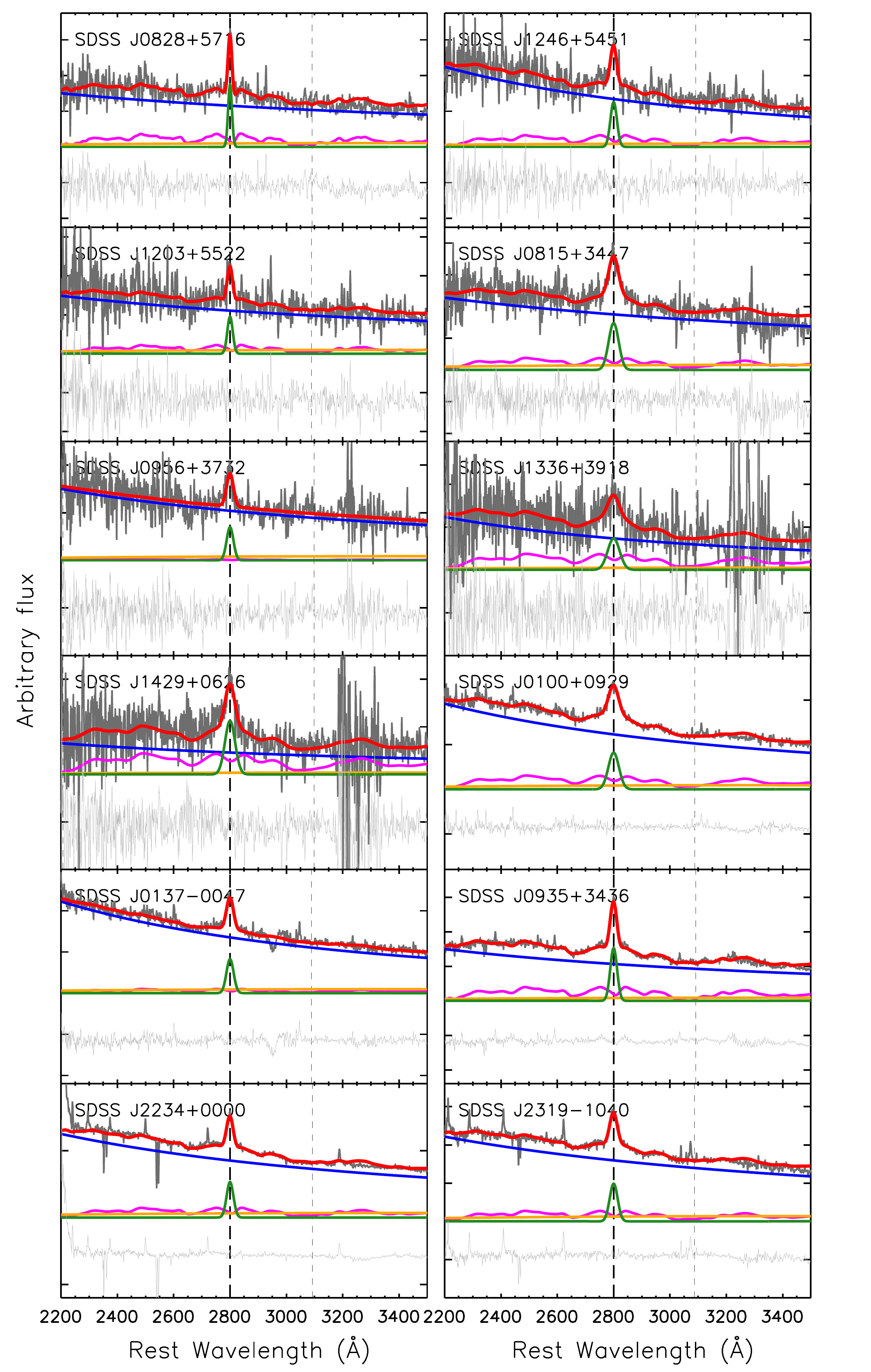}
\caption{
     Fully reduced GNIRS spectra of our targets in the spectral region around the \MgII\ emission
     after 3 pixel smoothing (gray solid line), where the best-fit results are overlaid. Total fit (red), 
     power-law continuum (blue), Balmer-pseudo continuum (orange), \FeII\ (magenta), and 
     \MgII\ (green) are shown respectively. The residual spectrum (thin gray solid line) is shown in 
     the lower panel for each target.
\label{fig:allspec1}}
\end{figure}

\subsection{\Hb\ and AGN properties} \label{section:Data}
We also measured AGN properties (i.e., $M_{\rm BH}$, luminosity, and Eddington ratio). 
The mass of SMBHs was calculated using the single-epoch method \citep{Vestergaard2002, 
Wu2004, Vestergaard2006, Park2012a,Matsuoka2013, Karouzos2015} based on the dispersion of 
emission lines and continuum luminosity. Specifically, three emission lines have been often used, 
namely, \Hb\ $\lambda$4861, \MgII, and \CIV, combined with the continuum luminosity 
density at 5100, 2800, and 1350\AA\ respectively. In this work, we used \Hb\ with the continuum 
luminosity density at 5100\AA\ for the high-z quasars since it is well calibrated compared 
to the other emission lines \citep[e.g.,][]{Park2012a}. For low-redshift quasars, we used \MgII\ 
and the continuum luminosity density at 3000\AA\ to measure the AGN properties since we have 
no \Hb\ information.

For fitting the spectral region around the \Hb\ emission, we took into account of a power-law 
continuum, \OIII, and narrow- and broad- components of \Hb. No apparent 
signature of the \FeII\ multiplet and host galaxy stellar absorption lines are present in the rest-frame
optical part of the spectra.   
Thus, we do not include the \FeII\ and stellar components in the fitting procedure. Note that 
it is a general trend that the \FeII\ multiplet emission in the rest-frame optical is weaker than that 
in the rest-frame UV \citep[e.g.,][]{Tsuzuki2006}. 

We fitted each of the \OIII $\lambda\lambda$4959,5007 doublet with a double Gaussian model
Similar to \MgII, the fit with the Gauss-Hermite series tends to be largely affected by noise and thus the 
Gauss-Hermite profile was not used for fitting the \OIII\ line. For the \Hb\ emission, we adopted 
the line profile of \OIII\ as the \Hb\ narrow component with a flux scaling, while we used a double Gaussian 
model to fit the broad \Hb\ line. Note that the narrow component of the \Hb\ to the \OIII\  flux ratio 
is within the range of 5--40\%, that is consistent with previous works \cite[e.g.,][]{Park2012a}. 
Then we calculated $M_{\rm BH}$ using the velocity dispersion of the broad \Hb\ broad line and 
the continuum luminosity density at 5100\AA\ by adopting Equation 2 of \cite{Woo2015}. 
Also, $M_{\rm BH}$ based on the velocity dispersion of \MgII\ and the continuum luminosity density at 3000\AA\ was calculated using the equation 
given in \citet{Woo2018}. The bolometric luminosity of AGNs was derived from the continuum luminosity density 
at 3000\AA\ and 5100\AA, by multiplying a bolometric correction factor of
5.15 (for 3000\AA) and 9.26 (for 5100\AA) given in \cite{Shen2008}.
The Eddington ratio was calculated by dividing the derived bolometric luminosity by the Eddington luminosity determined 
using $M_{\rm BH}$. The uncertainties of these AGN properties were estimated using  Monte-Carlo simulations. 
The results are listed in Table 2.

\begin{figure}
\centering
\includegraphics[height=15cm, width = 0.51\textwidth]{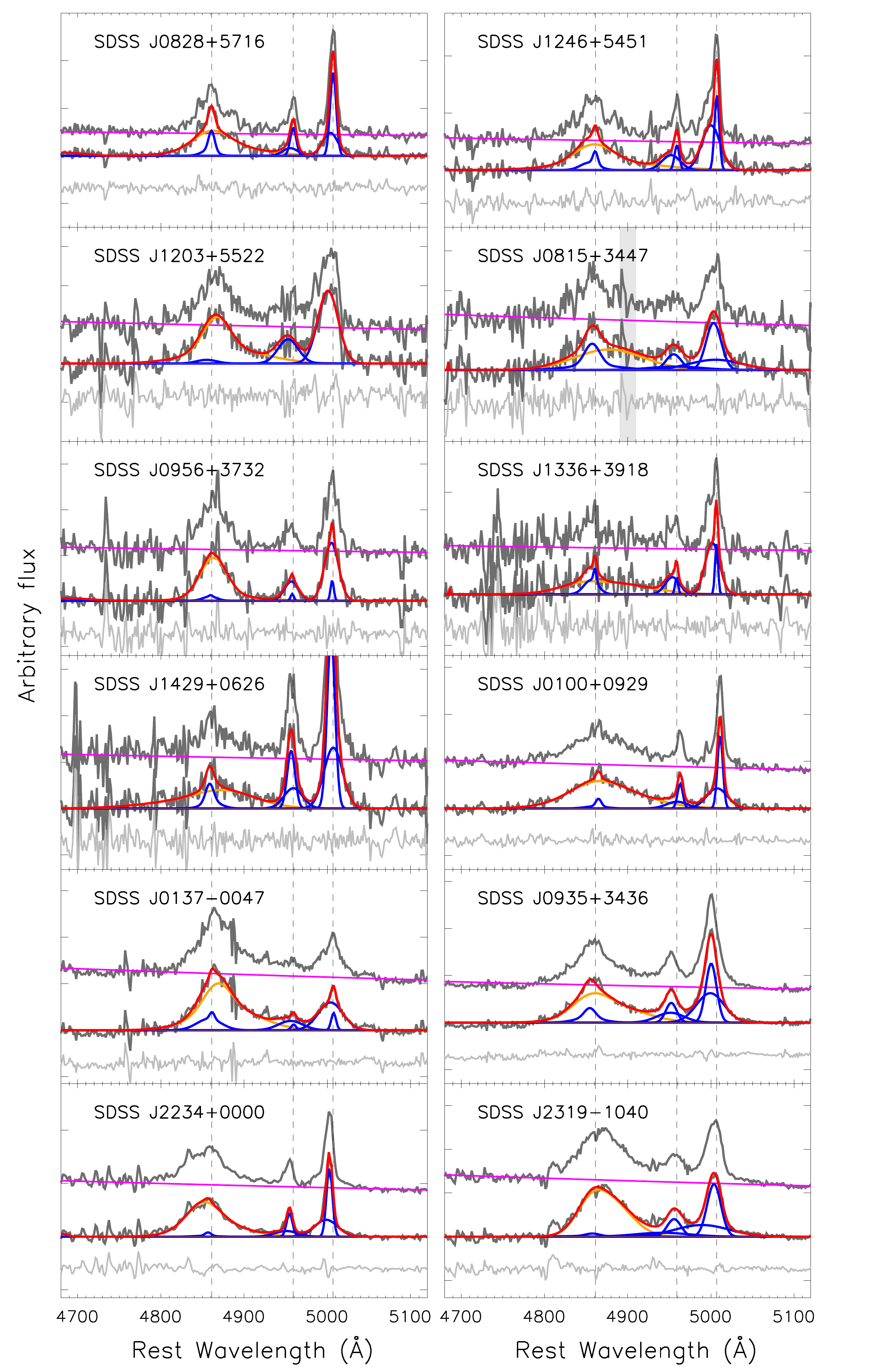}
\caption{
     Same as Figure 2 but for the spectral region around the \Hb\ emission. Total fit (red), 
     power-law continuum (magenta), and the \OIII\ and narrow \Hb\ components (blue) are 
     presented. Also, the broad component of \Hb\ is shown in orange.
\label{fig:allspec1}}
\end{figure}

\subsection{\NV/\CIV} \label{section:Data}
As an another indicator of chemical properties of the BLR in quasars, we measured the \NV/\CIV\ flux ratio 
of our sample using the optical SDSS 
spectra. Since many UV emission lines are heavily blended (e.g., \Lya 
$\lambda$1216 and \NV), we conducted multi-component fitting analysis by adopting the same
procedure as adopted by \citet{Shin2013}. In the fitting, we divided broad emission lines into two 
groups based on the ionization potential of the corresponding ion (i.e., high- and low-ionization 
emission lines) assuming the same kinematics (the line shift and 
dispersion) for these emission lines in each group \citep[see, e.g.,][]{McIntosh1999}. We adopted a double 
Gaussian model for fitting the emission lines. Then, the \NV/\CIV\ flux ratio was calculated from the 
best fit model. Its uncertainties were estimated from the signal-to-noise ratio of the SDSS 
spectrum, and the derived flux ratio and its uncertainty are presented in Table 2.

\begin{figure}
\centering{
\includegraphics[height=15cm, width = 0.51\textwidth]{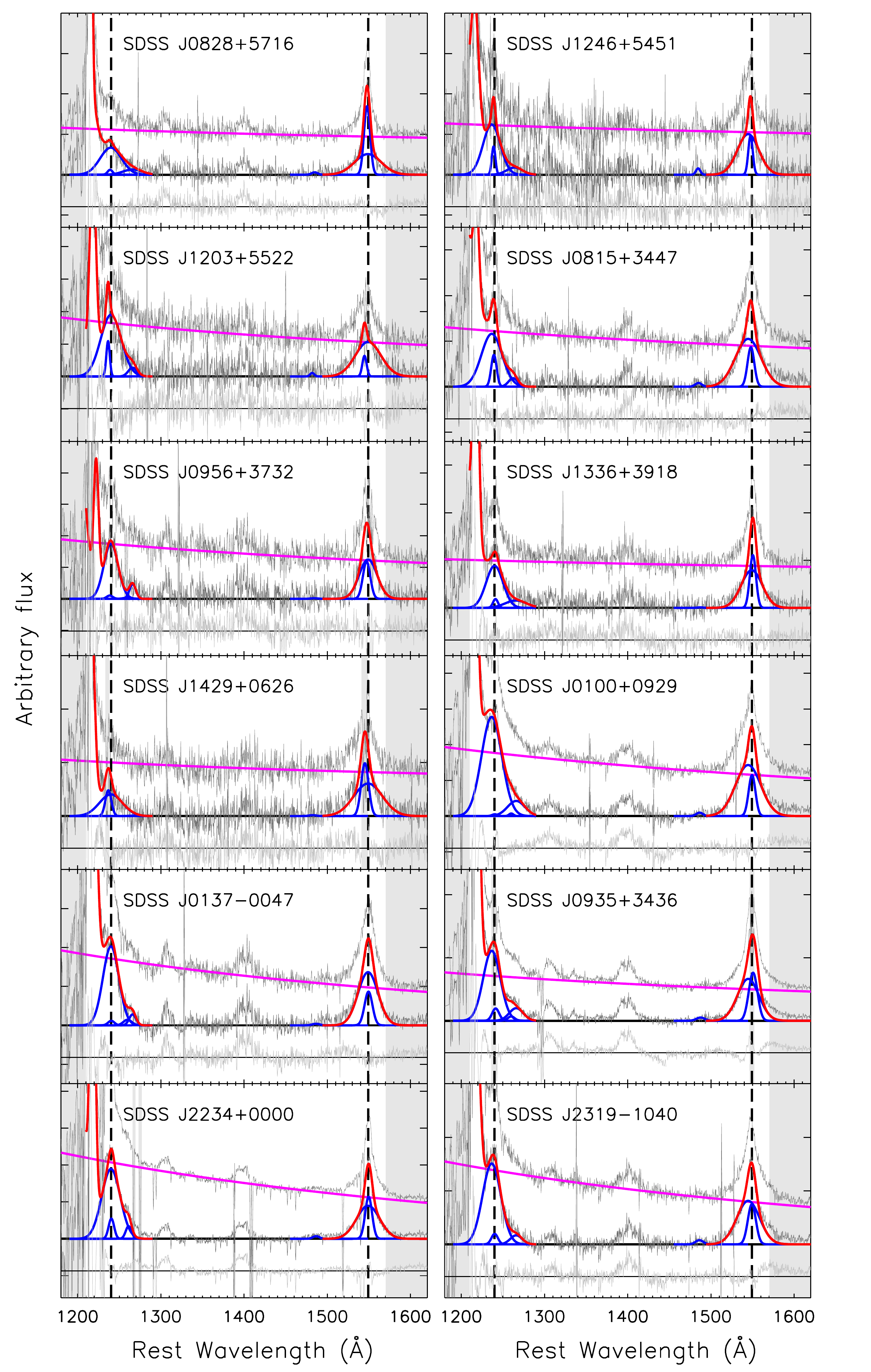}
\caption{
     SDSS spectra of our targets for the spectral region around the \NV\ and \CIV\ emission.
     Total fit (red), power-law continuum (magenta), and the \NV\ and \CIV\ components (blue) are 
     shown respectively. The residual spectrum (thin gray solid line) is shown in 
     the lower panel for each target.
     \label{fig:allspec1}}}
\end{figure}

\begin{deluxetable*}{cccccccccc} 
\tablewidth{0pt}
\tablecolumns{9}
\tabletypesize{\scriptsize}
\tablecaption{Line flux ratios and AGN properties}
\tablehead{
\colhead{Target} &
\colhead{log FeII/MgII} &
\colhead{log FeII/MgII} &
\colhead{log FeII/MgII} &
\colhead{log FeII/MgII} &
\colhead{log NV/CIV} &
\colhead{log $M_{\rm BH}/M_{\odot}$} &
\colhead{log $L_{\rm Bol}$} &
\colhead{log $L_{\rm Bol}/L_{\rm Edd}$} 
\\
 &
\colhead{(T06)} &
\colhead{(VW01)} &
\colhead{(mo\_VW01)} &
\colhead{(mo\_VW01$_{1860}$)} &
&
&
\colhead{(erg s$^{-1}$)}
&
\\
\colhead{(1)} &
\colhead{(2)} &
\colhead{(3)} &
\colhead{(4)} &
\colhead{(5)} &
\colhead{(6)} &
\colhead{(7)} &
\colhead{(8)} &
\colhead{(9)} }
\startdata
SDSS J082854.44+571637.2	&$	0.79	\pm	0.03	$&$	0.56	\pm	0.03	$&$	0.60	\pm	0.04	$&$	0.63	\pm	0.04	$&$	-0.21 \pm 0.10  $&$	9.08	\pm	0.27	$&$	46.62	\pm	0.00	$&$	-0.55	\pm	0.28	$\\
SDSS J124652.80+545140.6	&$	0.68	\pm	0.06	$&$	0.42	\pm	0.06	$&$	0.45	\pm	0.06	$&$	0.46	\pm	0.06	$&$	-0.04 \pm 0.15	$&$	9.15	\pm	0.22	$&$	46.65	\pm	0.00	$&$	-0.59	\pm	0.22	$\\
SDSS J120308.69+552245.8	&$	0.66	\pm	0.34	$&$	0.37	\pm	0.36	$&$	0.40	\pm	0.37	$&$	0.35	\pm	0.42	$&$	0.15 \pm 0.10 $&$	8.83	\pm	0.22	$&$	46.62	\pm	0.02	$&$	-0.30	\pm	0.23	$\\
SDSS J081528.12+344737.0	&$	0.47	\pm	0.24	$&$	0.27	\pm	0.20	$&$	0.28	\pm	0.21	$&$	0.30	\pm	0.20	$&$	-0.08 \pm 0.06	$&$	9.15	\pm	0.16	$&$	46.69	\pm	0.01	$&$	-0.56	\pm	0.17	$\\
SDSS J095617.14+373224.7	&$	<0.45^\ast		$&$	<-0.20^\ast	$&$	<-0.19^\ast	$&$	< -0.33^\ast	$&$	-0.08 \pm 0.09	$&$	8.39	\pm	0.31	$&$	46.39	\pm	0.01	$&$	-0.09	\pm	0.31	$\\
SDSS J133600.20+391826.2	&$	0.77	\pm	0.23	$&$	0.46	\pm	0.16	$&$	0.49	\pm	0.17	$&$	0.46	\pm	0.19	$&$	-0.17 \pm 0.12	$&$	8.91	\pm	0.31	$&$	46.20	\pm	0.01	$&$	-0.81	\pm	0.31	$\\
SDSS J142903.86+062620.4	&$	0.69	\pm	0.14	$&$	0.42	\pm	0.13	$&$	0.44	\pm	0.14	$&$	0.43	\pm	0.14	$&$	-0.31 \pm 0.16	$&$	9.06	\pm	0.32	$&$	46.38	\pm	0.02	$&$	-0.78	\pm	0.34	$\\
SDSS J010049.76+092936.1	&$	0.65	\pm	0.02	$&$	0.42	\pm	0.02	$&$	0.44	\pm	0.02	$&$	0.43	\pm	0.02	$&$	0.10 \pm 0.03	$&$	9.55	\pm	0.17	$&$	47.22	\pm	0.00	$&$	-0.42	\pm	0.17	$\\
SDSS J013735.46-004723.4	&$	0.31	\pm	0.06	$&$	0.04	\pm	0.07	$&$	0.06	\pm	0.07	$&$	0.05	\pm	0.08	$&$	-0.01 \pm 0.04	$&$	9.08	\pm	0.05	$&$	46.97	\pm	0.00	$&$	-0.21	\pm	0.05	$\\
SDSS J093514.41+343659.5	&$	0.66	\pm	0.01	$&$	0.44	\pm	0.01	$&$	0.47	\pm	0.01	$&$	0.47	\pm	0.01	$&$	-0.01 \pm 0.03	$&$	9.23	\pm	0.04	$&$	47.19	\pm	0.00	$&$	-0.14	\pm	0.05	$\\
SDSS J223408.99+000001.6	&$	0.68	\pm	0.01	$&$	0.43	\pm	0.01	$&$	0.45	\pm	0.01	$&$	0.46	\pm	0.01	$&$	0.12 \pm 0.01	$&$	9.42	\pm	0.01	$&$	47.82	\pm	0.00	$&$	0.29	\pm	0.01	$\\
SDSS J231934.77--104037.0	&$	0.63	\pm	0.02	$&$	0.40	\pm	0.01	$&$	0.42	\pm	0.01	$&$	0.42	\pm	0.02	$&$	0.07 \pm 0.03	$&$	9.08	\pm	0.09	$&$	47.39	\pm	0.00	$&$	0.21	\pm	0.09	$
\enddata
\label{table:prop}
\tablecomments{
    $\ast$: 2 sigma upper limit.}
\end{deluxetable*}

\begin{figure*}
\centering
\includegraphics[width = 0.8\textwidth]{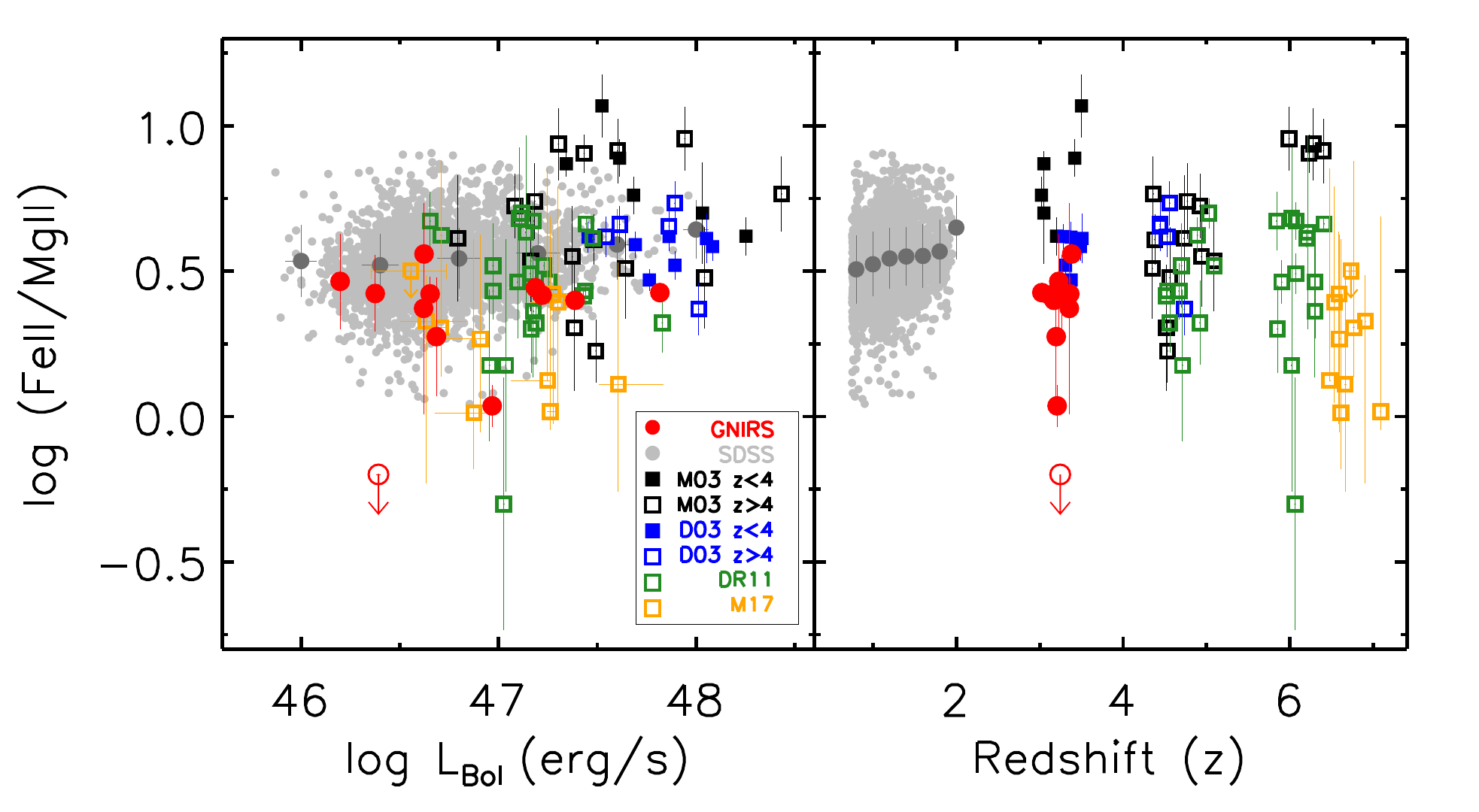}
\caption{
     \FeII/\MgII\ ratio as a function of AGN bolometric luminosity (left) and redshift (right).
     Our $z\sim3$ sample (red) is compared to the samples from 
     \citet[][M03]{Maiolino2003} (black),
     \citet[][D03]{Dietrich2003} (blue),
     \citet[][DR11]{DeRosa2011} (green), and \citet[][M17]{Mazzucchelli2017} (yellow). 
     Our low-z SDSS sample is denoted with light-gray (each object) and dark-gray circles (binned data).  
     Filled squares indicate the measurements for quasars at $z<4$, while open squares 
     show the measurements for high-redshift ($>4$) quasars. One AGN with an upper limit of \FeII/MgII\ in our sample is
     denoted with an open red circle. Note that for comparing with the \FeII/\MgII\ ratio of the literature samples, we adopted the \FeII/\MgII\ ratio
     measurements based on the VW01 template (see Table 2).
\label{fig:allspec1}}
\end{figure*}

\section{result} \label{section:result}

\subsection{\FeII/\MgII\ ratio}
In Figure~5, we compare the \FeII/\MgII\ flux ratio of our quasar sample with the previous results from the 
literature \citep{Maiolino2003,Dietrich2003,DeRosa2011,Mazzucchelli2017} as a function of the redshift and 
bolometric luminosity. We note that \cite{Maiolino2003} and \cite{Dietrich2003} provided only 
$L_{1450}$ and we converted it to bolometric luminosity using the bolometric correction factor of 
4.2 \citep{Runnoe2012}.
Here we present the \FeII/\MgII\ measurements based on the VW01 template in Figure~5 to be 
consistent with previous results \citep{Maiolino2003,Dietrich2003,DeRosa2011,Mazzucchelli2017}.
Note that the \FeII/\MgII\ flux ratios based on VW01 and mo\_VW01 (also 
mo\_VW01\_1860) are consistent within $\sim$0.025 dex ($\sim$ 6\%) on average, and 
thus, this difference does not change any trend in Figure~5.

We find that, among our sample at $z\sim 3$, lower-luminosity quasars do not show lower \FeII/\MgII\ flux ratios 
than higher-luminosity quasars. In other words, the \FeII/\MgII\ flux ratio appears to be independent 
of the AGN luminosity for our quasars at $z\sim 3$ as they show similar values within the
uncertainty except for two outliers (SDSS J0137--0047 and SDSS J0956+3732). In the case of
the SDSS quasars at $z \lesssim 2$, while bolometric luminosity is similar to that of our sample,
\FeII/\MgII\ flux ratio is similar to that of $z\sim 3$ quasars in our sample. 
We calculate the mean \FeII/\MgII\ flux ratio of the SDSS quasars in each luminosity/redshift 
bin and the uncertainties of them were taken from the standard deviation of the values in each bin. 
Even though the mean values of \FeII/\MgII\ show weak evolution as a function of redshift, they are 
consistent within the uncertainties.  
In contrast, our measurements of \FeII/\MgII\ flux ratio are significantly lower than those of 
\cite{Maiolino2003} at similar redshift ($z \sim 3$), indicating the effect of luminosity.  
We will discuss possible origins of this discrepancy in the next section.

We also find that there is no significant \FeII/\MgII\ evolution as a function of redshift. Even though 
there are systematic differences between each study, our result is consistent with previous 
works \citep[e.g.,][]{DeRosa2011,Mazzucchelli2017}. For assessing the statistical significance quantitatively, we 
performed Spearman's lank-order statistical test for the relation between the \FeII/\MgII\ flux ratio 
and the redshift, and also between the \FeII/\MgII\ flux ratio and $L_{\rm Bol}$. 
The obtained correlation coefficients and Spearman's lank-order probabilities for only high-z
quasars (except for low-redshift SDSS quasars and SDSS J0956+3732) are 0.327 and 0.007 for 
$L_{\rm Bol}$, and 0.125 and 0.313 for the redshift. If we include low-z SDSS quasars, the 
coefficients become 0.166 and 3.87$\times 10^{-27}$ for $L_{\rm Bol}$, and 0.167 and 1.95 $\times 
10^{-27}$ for the redshift. The results of Spearman's lank-order statistical test infer a positive 
correlation between the \FeII/\MgII\ flux ratio and both redshift and luminosity, if quasars at all
redshift ranges are considered, but this result is presumably mainly due to the objects with high-luminosity 
and high \FeII/MgII\ ratio from \cite{Maiolino2003}. 
Since there is a large scatter of \FeII/\MgII\ flux ratio in each sample and in the total sample, 
it is difficult to confirm the correlation.
 

%

\begin{figure*}
\centering
\includegraphics[width = 0.9\textwidth]{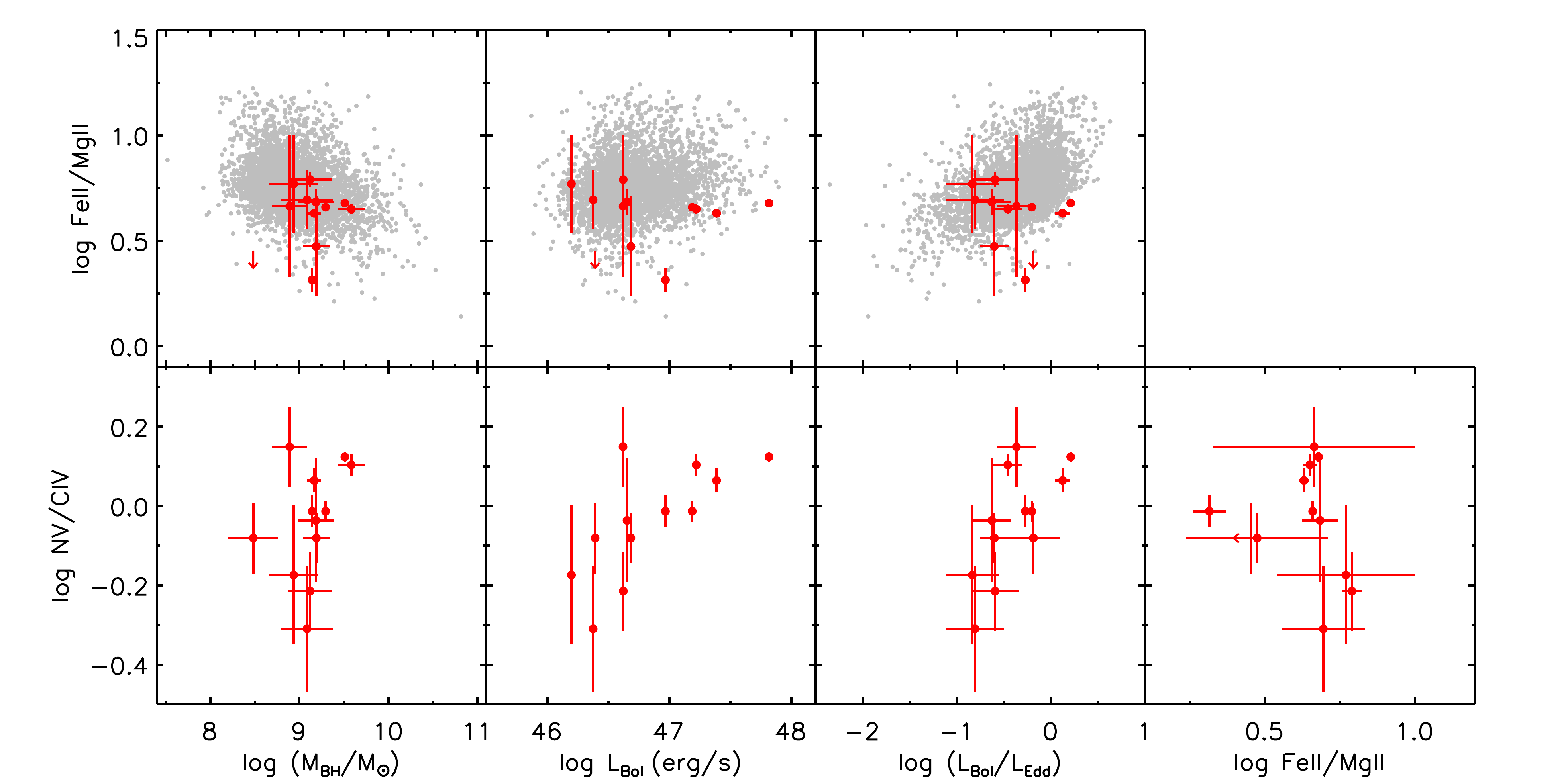}
\caption{
    Relation between two BLR emission-line flux ratios (\FeII/\MgII\ and \NV/\CIV\ in upper and lower panels
    respectively) and AGN properties (black hole mass, bolometric luminosity, and Eddington ratio,
    shown in left, middle, and right panels, respectively).
    Filled red and gray circles denote high-$z$ and low-$z$ samples of ours, respectively.} 
\end{figure*}

\subsection{The metallicity dependence on AGN properties}\label{section: comparison} 
We investigate the dependency of the \FeII/\MgII\ (based on the T06 template) and \NV/\CIV\ 
flux ratios on the AGN properties, i.e., the black hole mass, bolometric luminosity, and 
Eddington ratio. In Figure~6, the relations between \FeII/\MgII\ and AGN properties 
of low-z quasars show similar trend with those in \citet{Dong2011} and \citet{Sameshima2017}.
In contrast, the \FeII/\MgII\ of our sample of 12 high-z quasars show no clear dependency on AGN properties, presumably due to the small sample
size, while their flux ratios are consistent with those of low-z AGNs. 
Spearman's correlation test confirms a strong correlation between \FeII/\MgII\ and AGN properties for the combined sample of low-redshift SDSS 
quasars and high-redshift quasars, while little correlation is found between AGN properties and \FeII/\MgII\ if we use only high-redshift quasars (see Table 3).
In the case of the \NV/\CIV\ flux ratio, we find strong dependence on the
AGN properties, i.e., luminosity, mass, and Eddington ratio \citep[see, e.g.,][]{Matsuoka2011b}. We also investigate the relation between \FeII/\MgII\ and \NV/\CIV,
finding no clear correlation. As suggested for the relation between \FeII/\MgII\ and AGN properties in our sample, 
this non-correlation is presumably due to the limited sample size.
\\


\begin{deluxetable}{cccc} 
\tablewidth{0pt}
\tablecolumns{8}
\tabletypesize{\scriptsize}
\tablecaption{Results of Spearman's Rank-order correlation test}
\tablehead{
&
\colhead{$M_{\rm BH}/M_{\odot}$} &
\colhead{$L_{\rm Bol}/L_\odot$} &
\colhead{$L_{\rm Bol}/L_{\rm Edd}$} 
\\
\colhead{(1)} &
\colhead{(2)} &
\colhead{(3)} &
\colhead{(4)}  }
\startdata
\FeII/\MgII	&$r_{s}$= --0.409 & $r_{s}$= --0.582 & $r_{s}$= --0.500 	\\
		&$p$=  0.212  &$p$= 0.060 &$p$= 0.117         \\
\FeII/\MgII\ w/ SDSS	&$r_{s}$= --0.309 & $r_{s}$= 0.189 & $r_{s}$= 0.486 	\\
		&$p$= 0.000 &$p$= 0.000 &$p$= 0.000         \\
\NV/\CIV	&$r_{s}$= 0.400 & $r_{s}$= 0.670 & $r_{s}$= 0.642	\\
		&$p$= 0.197 &$p$= 0.017 &$p$=      0.024
\enddata
\label{table:prop}
\end{deluxetable}

\section{Discussion}\label{section:Discussion}
\subsection{No \FeII/\MgII\ evolution of BLR in quasars}\label{starformation}
As discussed in Section~1, it is expected that low-luminosity quasars at high-z may show 
signatures of variations in the chemical composition traced by the \FeII/\MgII\ flux ratio,  
compared to low-z quasars \citep{Kawakatu2003,Juarez2009}.
However, the observed \FeII/\MgII\ ratio of low-luminosity quasars (log $L_{\rm bol} \sim 46.5$) at z$\sim 3$
is comparable to those of low-z quasars and those of high-luminosity quasars (log $L_{\rm bol} \sim 47.5$) at similar redshift.
Also, the \FeII/\MgII\ ratios in high-$z$ quasars (regardless of
the bolometric luminosity) are similar to those in low-$z$ quasars.
This suggests that there is no significant redshift evolution in the Fe/Mg abundance ratio. 
Though the \NV/\CIV\ flux ratio in our 
high-$z$ quasar sample show a clear dependence on the bolometric luminosity, 
the dependence does not show any redshift evolution (see \citealt{Nagao2006, Matsuoka2011b}).
Therefore, we conclude that there is no chemical evolution including the Fe/Mg
abundance ratio, in quasars for a wide redshift range, i.e., $0 \lesssim z \lesssim 3$. In other words,
observed quasars are already matured chemically, even for high-$z$ low-luminosity quasars. 

There are two exceptions showing a low \FeII/\MgII\ flux ratio (SDSS J0137--0047 and SDSS 
J0956+3732), which may imply that they are in a chemically young phase. However, 
since one of the two quasars, SDSS J0137--0047, has relatively high luminosity ($L_{\rm bol} = 
47.0$ erg s$^{-1}$), it does not support the idea that low-luminosity quasars are in a 
chemically young phase. One possible interpretation for the no evolution of \FeII/\MgII\ flux ratio is
that our targets are still too luminous to look at quasars in a chemically-young phase.
\cite{Maiolino2008} showed that the metallicity evolution of massive galaxies ($M_{\ast}/M_{\odot} 
\sim\ 10^{11}$) is less prominent than that of low-mass galaxies $M_{\ast}/M_{\odot} < 10^{10}$ 
for $0 < z < 3$. Since the black hole mass of our high-$z$ quasar sample is typically $\sim$10$^9$
$M_\odot$ (Table 2), the mass of their host galaxies is inferred to be $\sim$10$^{11-12}\ M_{\odot}$ by 
assuming the \mbh-to-stellar mass ratio to be $0.001-0.01$\footnote{
The mass ratio of SMBHs to their host galaxies (specifically their bulge component) is 
$\sim 0.002$ (e.g., \citealt{Marconi2003}) in the nearby Universe. The mass ratio at high-$z$
is not clearly measured, and it may increase up to $\sim 0.01$ at $z \sim 3$ (e.g., \citealt{Peng2006,Schramm2008}; 
but see also \citealt{Schulze2014}).} 
Thus, our results suggest that our targets are already chemically evolved with the previous type 1a SN activities. To search for 
chemically-young quasars, it is necessary to use lower-luminosity quasars, whose \mbh\ is expected to be on average lower. 

\subsection{\FeII/\MgII\ and AGN properties}\label{starformation}

In Figure~6, the \FeII/\MgII\ ratios of low-redshift quasars show negative (positive) correlation with 
black hole mass (Eddington ratio) as reported by \citet{Dong2011}. Since 
galaxy mass correlates with the Mg/Fe abundance ratio \citep[e.g.,][]{Thomas2005,Johansson2012,Conroy2014} 
as well as galaxy black hole mass \citep[e.g.,][]{Magorrian1998,Ferrarese2000,Gebhardt2000,
Marconi2003,Haring2004,Kormendy2013,Woo2013,Woo2015}, the correlation between 
the \FeII/\MgII\ flux ratio and \mbh\ is naturally expected, if the \FeII/\MgII\ flux ratio is a good
indicator of the Fe/Mg abundance ratio.

On the other hand, \cite{Dong2011} also reported that the \FeII/\MgII\ flux ratio shows a stronger 
dependency on the Eddington ratio than on the black hole mass, suggesting that the \FeII/\MgII\ flux ratio is more likely 
governed by the gas density of the BLR, rather than the Fe/Mg abundance ratio \citep[see also][]{Sameshima2017}. 
Using the photoionization code, Cloudy \citep{Ferland1998}, several theoretical works also showed that 
the \FeII/\MgII\ flux ratio depends on various physical parameters such as the turbulence and the BLR gas density \citep[e.g.,][]{Verner2003,Baldwin2004,Sameshima2017}. 
To investigate the \FeII/\MgII\ correlation with Eddington ratio, \cite{Sameshima2017} hypothesized that gas density correlates with 
Eddington ratio and discussed the possibility of the BLR radius-Eddington ratio relation \citep{Rees1989}, while
the origin of gas density-Eddington ratio relation has not been discussed. 
By proposing the correction for the gas density dependence, they discussed 
the intrinsic Fe/Mg abundance ratio. 
Note that using the correction method (Eq. 12 of \citealt{Sameshima2017}), we find only 0.02 dex difference in the 
\FeII/\MgII\ flux ratio for our high-z quasars.

While the 12 high-z quasars in our sample have similar AGN properties compare to the low-z quasars,
we do not find clear relations with AGN properties, presumably due to the relatively large error of the \FeII/\MgII\ flux ratio
and the small sample size. To better understand these relations for high-z quasars, a large spectroscopic sample is needed.

\begin{figure*}
\centering
\includegraphics[width = 0.9\textwidth]{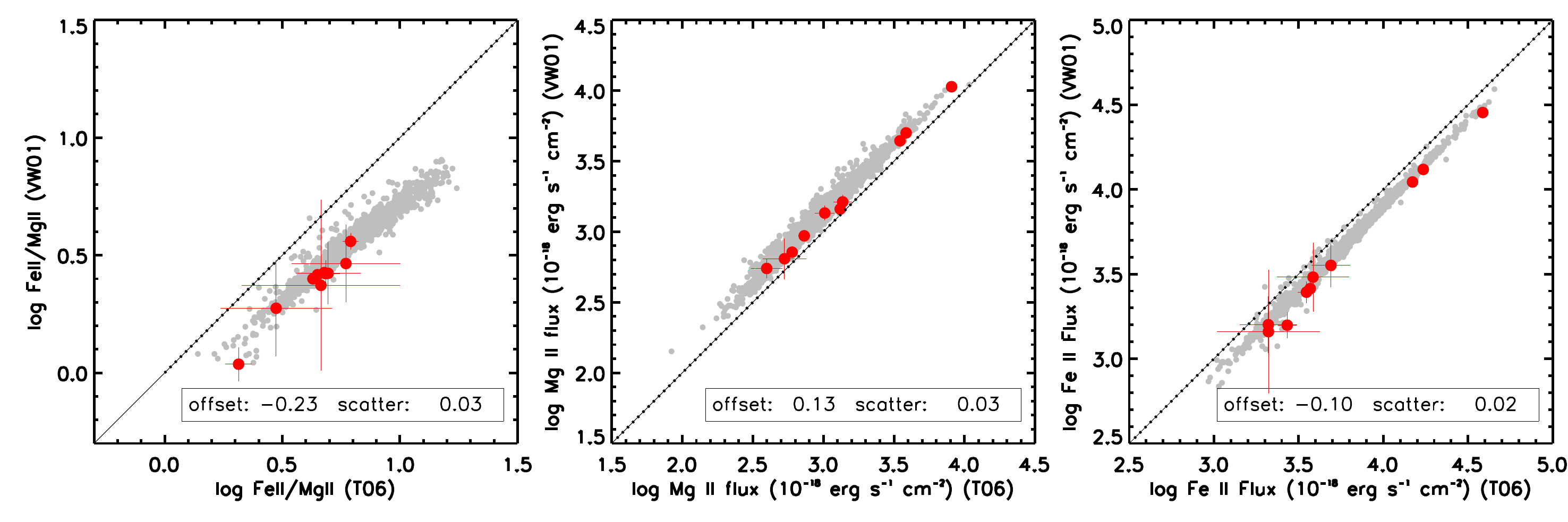}
\caption{
     Comparison of the measurement results by adopting different iron templates. 
     Measurements with the T06 template and the VW01 template are shown in X-axis and Y-axis,
     respectively. Symbols are the same as in Figure 5.}
\end{figure*}

\begin{figure*}
\centering
\includegraphics[width = 0.9\textwidth]{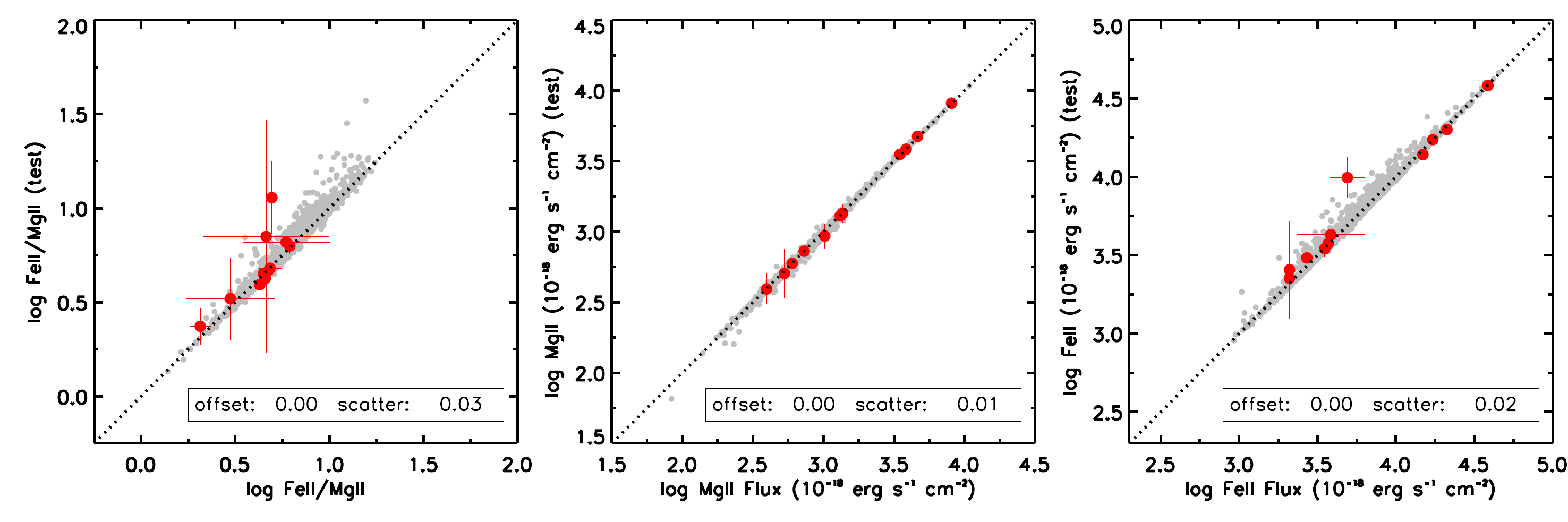}
\caption{
  Comparison of the measurement results with/without velocity constraint. Original 
     measurements (with velocity constraint) presented in Table~2 are shown in X-axis, while the results without velocity 
     constraint are shown in Y-axis. Symbols are the same as in Figure 5. }
\end{figure*}

\subsection{Systematic uncertainties of the \FeII/\MgII\ flux ratio}\label{section: RD}

There is a discrepancy of the \FeII/\MgII\ flux ratio measurements among various studies as 
shown in Figure~5. For example, for one target in our sample SDSS J2234+0000,
\citet{Maiolino2003} reported the \FeII/\MgII\ ratio as $5.77\pm0.86$ while 
our measurement is $2.68\pm0.06$, showing a factor of $\sim 2$ difference.
\cite{Kurk2007} discussed several factors (e.g., the adopted iron template 
and the fitting method) that can severely affect the measurement of the \FeII/\MgII\ flux ratio,
and found that such effects can result in a large difference in the measured \FeII/\MgII\ flux ratio
by a factor of up to 2. 

In order to investigate the systematic uncertainty in the \FeII/\MgII\ measurement, 
we conducted a couple of tests using the combined sample of 11 high-z quasars and $\sim$4000 low-z quasars. 
Note that we exclude SDSS J0956+3732 since its \FeII\ emission
is not clearly detected in our observation (Section 4.1).
First, we investigate the effect of the iron templates.
As shown in Figure~1, the T06 template successfully recovers \FeII\ around the \MgII\ emission 
at 2800\AA, which is not recovered by the VW01 template. We compare the \FeII/\MgII\ flux ratios along with 
the fluxes of \MgII, and \FeII, based on the templates of VW01 or T06 in Figure~7. We find 
that the T06 template provides an average higher \MgII\ flux (0.13 dex), lower \FeII\ flux (--0.10 dex), and higher 
\FeII/\MgII\ flux ratio (0.23 dex) than the VW01 template. 
This result indicates that the \FeII/\MgII\ flux ratio is underestimated when the VW01 template is adopted. 
If we use the modified VW01 template or the modified VW01 template with a fixed velocity dispersion,
the \FeII/\MgII\ measurements are similar to those based on the VW01 template. 

Second, we test the effect caused various fitting constraints, using the T06 template for the fitting. 
As shown in Figure~2, the \MgII\ emission is clearly present in the spectra 
while the excess of the \FeII\ emission above the continuum is not very significant in many 
cases. Therefore the measurement of the \FeII\ flux can be sensitive to the details of the fitting method. 
As described in Section 3.1, we used the full wavelength range of 2200--3090\AA\ for 
the \FeII\ fitting with a constraint of the same velocity dispersion between the \MgII\ and \FeII\ emission. 
We test the \FeII/\MgII\ measurements with (1) no velocity dispersion constraint, (2) a narrower fitting range, and (3) 
a degraded spectral resolution as shown in Figures~7--9. 

While \FeII\ and \MgII\ have similar ionization energies, using the same velocity dispersion between them may 
introduce a systematic uncertainty if the kinematics of the two ions are different.
To examine this effect, we fit the spectra without tying the velocity dispersion between the \FeII\ and \MgII\ emission.
We find that  the \FeII/\MgII\ flux ratio with and without the velocity dispersion constraint are consistent to each other 
within 7\%, implying that the constraint on the velocity dispersion produces no significant difference (see Figure 7).

For investigating the effect of the fitting range, we fit the spectra with a narrower fitting 
range of 2600--3050\AA\, instead of the range of 2200--3090\AA\ (Figure~9). We find that the measurements
from the two fitting ranges are consistent within 0.02 dex ($5\%$), with a scatter of 0.09 dex ($23\%$). 
The scatter is mainly due to the measurements of \FeII, rather than those of \MgII\ (see the middle and right panels in Figure~9).
It is possible that the iron model is not perfect over the large fitting range. Thus, the measurements
vary depending on the fitting range. Data quality is another cause of the scatter since there is an offset of 0.02 dex for the low-flux subsample 
(i.e., $<$ $10^{-14} \rm \ erg\ s^{-1}\ cm^{-2}$) while the higher-flux quasars show no offset (see the right panel in Figure~9). 

We then test the effect of the spectral resolution. In Figure~5, we show that the measured
\FeII/\MgII\ flux ratios by \cite{Maiolino2003} are systematically higher those in other studies 
(\citealt{Dietrich2003} and this work). \cite{Maiolino2003} used the low spectral resolution, 
which was R$\sim$75 corresponding to 4000 km s$^{-1}$ in the velocity
resolution. This low resolution may be a source of systematic uncertainties of the \FeII/
\MgII. To examine the spectral resolution effect, we smooth our spectra down 
to $R \sim 75$ and measure the \FeII/\MgII\ flux ratio in the smoothed spectra. The measured 
\FeII/\MgII\ ratios before and after the smoothing are compared in Figure~10, which shows 
 a 0.09 dex scatter.

In summary, we find that adopting a proper iron template is the most important in measuring \FeII\ and \MgII\
fluxes. While the velocity dispersion constraint no significantly affect the \FeII\ and \MgII\ flux measurement,
the fitting range and the spectral resolution introduces non-negligible uncertainties. 
For the investigation of the \FeII/\MgII\ flux ratio, a sufficiently large sample is necessary  
to overcome the systematic uncertainties and avoid the sample selection bias.\\

\begin{figure*}
\centering
\includegraphics[width = 0.9\textwidth]{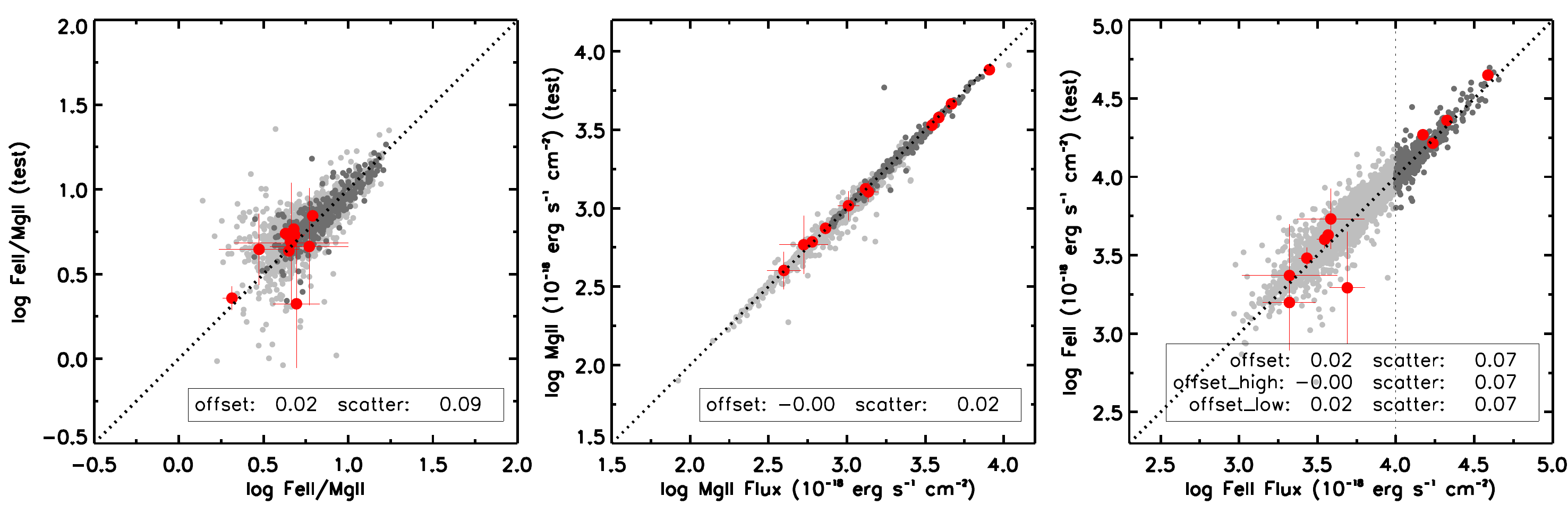}
\caption{
     Comparison of the measurement results by adopting two different fitting ranges. Our original 
     measurements based on the fit for the range of 2200--3090\AA\ and the measurements with a 
     narrower fitting range of 2600--3050\AA\ are shown in X- and Y- axes, respectively. 
     Filled red and light gray circles denote high-$z$ and low-$z$ samples of ours, respectively. Dark gray 
     circles denote low-$z$ sample with high flux ($\rm {\it F}_{\FeII}$ > $10^{-14}\ \rm erg\ s^{-1}\ cm^{-2}$).
     }
\end{figure*} 

 \begin{figure*}
\centering
\includegraphics[width = 0.9\textwidth]{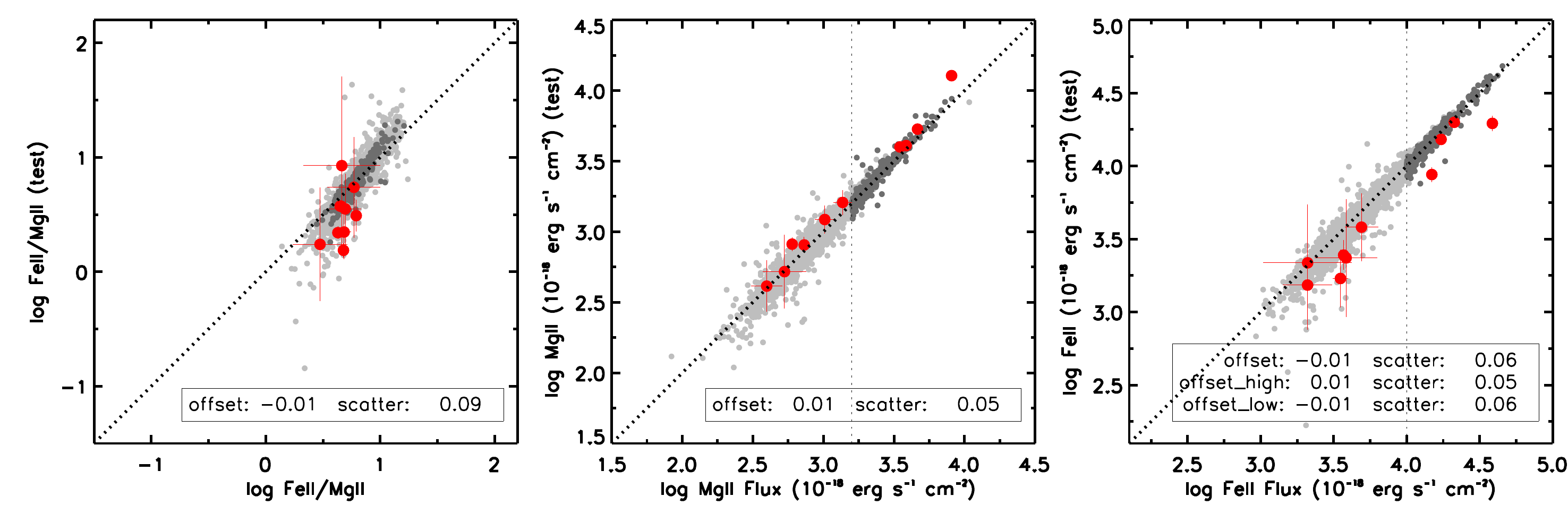}
\caption{
     Comparison between fittings with different resolution. 
     Our original measurements and measurements from the spectrum with $R\sim$75 are shown in X- and Y- axes.
     Filled red and light gray circles denote high-$z$ and low-$z$ samples of ours, respectively. 
     Dark gray circles shows low-$z$ sample with high flux ($\rm {\it F}_{\FeII}$ > $10^{-14}\ \rm \ erg\ s^{-1}\ cm^{-2}$ and 
     $\rm {\it F}_{\MgII}$ > $10^{-14.8}\ \rm \ erg\ s^{-1}\ cm^{-2}$).
}
\end{figure*} 


\section{Summary}\label{summary}
We present the analysis of the \FeII/\MgII\ flux ratio measurements for a sample of quasars at z$\sim 3$,
based on the NIR spectrum obtained with the Gemini/GNIRS.
The sample with an order of magnitude lower luminosity than the quasars at similar redshift studied in the previous
works, is selected to investigate  the evolution of \FeII/\MgII\ at high redshift $z\sim$3--6, since the low-luminosity
quasars may show a signature of chemically young status. 
We find that the low-luminosity quasars (log $L_{\rm bol} \sim 46.5$) show similar \FeII/\MgII\ ratios compared to those of high-luminosity quasars (log $L_{\rm bol} \sim 47.5$) at similar redshift, implying that even the low-luminosity quasars in our sample are already matured. 
To search for chemically young quasars with low \FeII/MgII\ ratios, quasars with much lower luminosity (i.e., log $L_{\rm bol} < 46 $) are necessary to study with upcoming extremely large telescopes.
We find that the \FeII/\MgII\ flux ratio correlates with black hole mass and Eddington ratio, which is consistent with the previous results in the literature.
By testing with various iron templates, fitting parameters, fitting range, we investigated systematic uncertainties,
concluding that a consistent method for measuring the \FeII/\MgII\ flux ratio is required to
compare with other works in investigating the evolution of the ratio. 

\acknowledgements
We thank the anonymous referee for useful suggestions. 
This work has been supported by the Basic Science Research Program through 
the National Research Foundation of Korea government (2016R1A2B3011457 and No.2017R1A5A1070354). 
TN is financially supported by JSPS (Grant 16H03958 and 17H01114).
The observations were carried out as a Subaru-Gemini time exchange program (ID: 2012A-C-003, PI: T. Nagao), where the travel expense was supported by the Subaru Telescope, which is operated by the National Astronomical Observatory of Japan (NAOJ), and also as a K-GMT science program (ID: 2015A-Q-203 \& ID: 2017B-Q-53, PI. J. Shin) of Korea Astronomy and Space Science Institute (KASI). We thank the staffs in the Gemini observatory for supporting our observations. 
Funding for the SDSS has been provided by the Alfred P. Sloan Foundation, the Participating Institutions, the National Science Foundation, the U.S. Department of Energy, the National Aeronautics and Space Administration, the Japanese Monbukagakusho, the Max Planck Society, and the Higher Education Funding Council for England. The SDSS Web Site is http://www.sdss.org/. The SDSS is managed by the Astrophysical Research Consortium for the Participating Institutions. The Participating Institutions are the American Museum of Natural History, Astrophysical Institute Potsdam, University of Basel, University of Cambridge, Case Western Reserve University, University of Chicago, Drexel University, Fermilab, the Institute for Advanced Study, the Japan Participation Group, Johns Hopkins University, the Joint Institute for Nuclear Astrophysics, the Kavli Institute for Particle Astrophysics and Cosmology, the Korean Scientist Group, the Chinese Academy of Sciences (LAMOST), Los Alamos National Laboratory, the Max-Planck-Institute for Astronomy (MPIA), the Max-Planck-Institute for Astrophysics (MPA), New Mexico State University, Ohio State University, University of Pittsburgh, University of Portsmouth, Princeton University, the United States Naval Observatory, and the University of Washington.


\end{document}